# ELECTRON CLOUD OBSERVATIONS AT THE ISIS PROTON SYNCHROTRON

A. Pertica (alex.pertica@stfc.ac.uk) and S. J. Payne (steve.payne@stfc.ac.uk)

ISIS Diagnostics Section, Rutherford Appleton Laboratory

Chilton, Didcot, Oxfordshire, OX11 0QX, UK

*Abstract*

The build up of electron clouds inside a particle accelerator vacuum chamber can produce strong transverse and longitudinal beam instabilities which in turn can lead to high levels of beam loss often requiring the accelerator to be run below its design specification.

To study the behaviour of electron clouds at the ISIS Proton Synchrotron, a Micro-Channel Plate (MCP) based electron cloud detector has been developed. The detector is based on the Retarding Field Analyser (RFA) design and consists of a retarding grid, which allows energy analysis of the electron signal, and a MCP assembly placed in front of the collector plate. The MCP assembly provides a current gain over the range 300 to 25K, thereby increasing the signal to noise ratio and dynamic range of the measurements.

This paper presents the first electron cloud observations at the ISIS Proton Synchrotron. These results are compared against signals from a beam position monitor and a fast beam loss monitor installed at the same location.

## INTRODUCTION

ISIS is the UK's spallation neutron and muon source located at the Rutherford Appleton Laboratory in Oxfordshire. The ISIS machine consists of a 50Hz proton synchrotron that accelerates ~3E13 protons per pulse from the Linac output energy of 70Mev to an extraction energy to target of 800MeV.

Electrons clouds are mainly seeded from electrons that are liberated though interactions of the particle beam with the residual gas, and from interactions of the beam with the vacuum chamber walls [1]. If the electron cloud density becomes large enough, this will lead to a strong transverse mode coupling between the proton beam and the electrons, in turn leading to beam instability and beam emittance growth and subsequent beam loss [2] [3].

To date there has never been any evidence, in the ISIS accelerator, of beam instabilities that have been directly associated with the production of electron clouds. Theoretical studies of the electron cloud phenomena based on the present ISIS accelerator and different ISIS upgrades [4] have indicated that with higher beam intensities, electron clouds could have an adverse effect on the performance of the ISIS accelerator [5].

In line with the proposed ISIS mega-watt upgrade plans [6] a decision was made to design and build detectors that could look for signs of electron cloud build-up and beam instabilities.

Two Retarding Field Analyser (RFA) based detectors were designed and installed into the 800MeV accelerator ring [7]. Results from these two RFA detectors showed no clear evidence of electron cloud activity. At the time this lack of visible electron cloud signal was considered to be due to the low level of the electron cloud signal combined with the high level of beam induced noise picked up by the RFA detectors themselves.

In order to overcome these two challenges, a Micro-Channel Plate (MCP) based electron cloud detector was developed. The detector is of a standard RFA design and a MCP assembly placed in front of a graphite collector plate. The MCP assembly provides an extra current gain, thereby increasing the signal to noise ratio and dynamic range of the electron cloud measurements.

## DETECTOR SETUP

*Detector assembly*

The new ISIS Electron Cloud Monitor detector assembly is shown in figure 1.

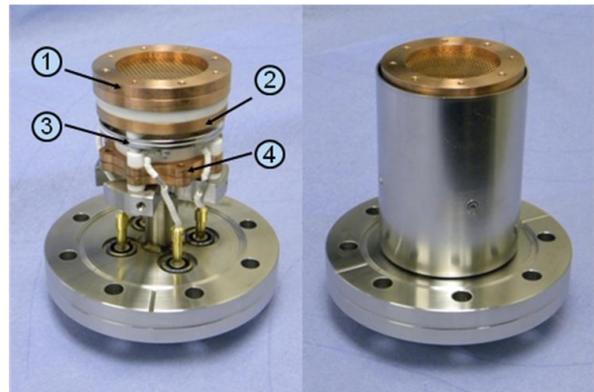

Figure 1: Details of the ISIS Electron Cloud Monitor. On the right hand side, the detector is shown with the mu metal shield in position.

At the front of the detector shown in figure 1 there is a 40mm diameter grounded grid (1) with a transmission efficiency of 20%. This grid acts as a substitute for the vacuum chamber wall so as to not interfere with the multipacting process [8], while at the same time allowing through a sample of electron flux from the cloud into the detector. Below the front grid, there is a 40mm diameter retarding grid (2) with a transmission efficiency of 85% connected to a negative power supply whose voltage output can be adjusted remotely, allowing measurements of the energy spectrum of the incoming electrons. The MCP assembly (3) is located between the retarding grid and the collector plate (4). The front plate of the MCP is grounded to the body of the detector and the back plate is connected to a positive bias supply. The collector assembly consists of a 55 mm diameter graphite disk sandwiched inside a copper frame. Graphite was chosen as the collector material due to its low secondary emission yield. The whole detector is surrounded by a low mu metal shield grounded to reduce the electromagnetic interference.

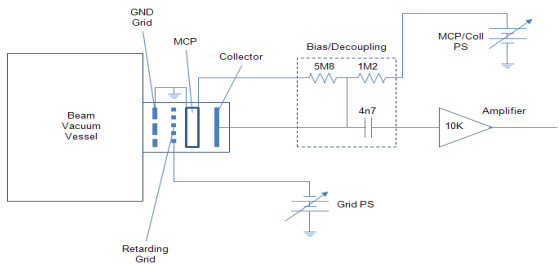

Figure 2: A schematic diagram of the detector setup.

## Amplifier

The amplifier used is a custom design d.c. coupled trans-impedance amplifier with a gain of 10 Kohms and a bandwidth of 40 MHz. The input impedance is 100 ohms and the output impedance 50 ohms. Because of the input a.c. coupling capacitor, which removes the collector/MCP d.c. bias voltage (Figure 2), the whole system has a band-pass response with cut-off frequencies of 330 KHz and 40 MHz.

## Location

The electron cloud monitor is positioned on the inside of the accelerator ring looking along the horizontal plane of the beam (Figure 3). At the same location there are the two RFA type detectors without MCP assemblies. One of these RFA's is located on the opposite side to the MCP based version, again looking along the horizontal plane of the beam. The second non-MCP RFA detector is situated on top of the beam pipe looking down onto the beam path.

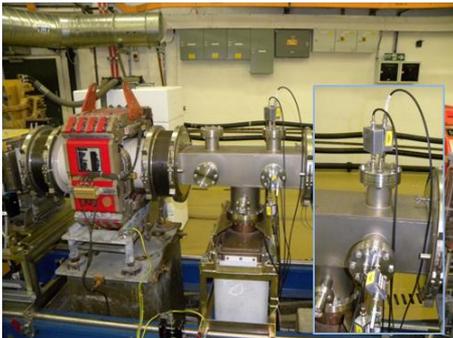

Figure 3: Detail of the installed detectors at the ISIS straight 5 location.

The detectors are located within a drift space of a straight section of the synchrotron. This location was chosen mainly because the space was available in that area of the ring.

## PRELIMINARY TESTS

### MCP gain

To match the electron signal amplitude to the input range of the amplifier/ADC setup, the MCP gain can be adjusted by varying the bias voltage applied to the MCP. If the input signal is not large enough compared to the background noise, the MCP gain can be increased to allow the use of the whole ADC range and increase the signal to noise ratio. If the input signal is too large, the MCP gain can be decreased in order to avoid saturation of the amplifier/ADC. During the tests, a range of bias voltages were chosen and the different waveforms were recorded. These measurements were carried out with a stable machine setup and applying averaging to the signals. Data taken after the injection stage, at different MCP bias voltages with the corresponding amplitude variations, is shown on figure 4.

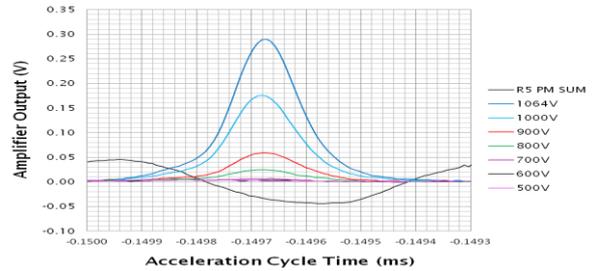

Figure 4: MCP gain test. EC signals corresponding to different MCP bias voltages applied.

### Retarding grid

By applying different negative voltages to the retarding grid, it should be possible to filter out those electrons with energies less than the grid potential. The following graph (Figure 5) shows the magnitude of the electron signal as a function of the grid voltage over the range 0V to 300V. As expected, the results show a reduction of the electron signal reaching the collector as the grid potential becomes more negative.

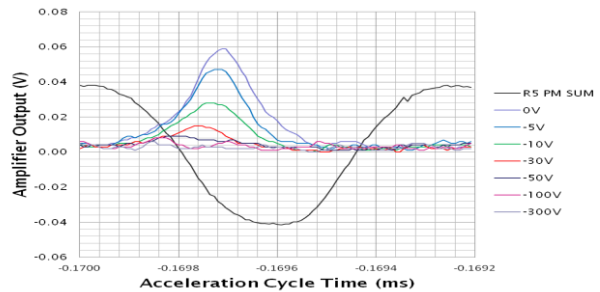

Figure 5: Retarding grid test. EC signals measured at different grid potentials.

# EXPERIMENTAL RESULTS

## Electron cloud observations

During the ISIS acceleration cycle, the largest electron cloud signals have been observed just after the injection stage. This is where the largest beam loses occur. Conversely the electron signal is smallest at beam extraction (10ms after injection), and this is where the beam induced electrical noise is quite large, so averaging and background removal techniques have been applied in order to obtain a clearer signal. The figures 6, 7 and 8 show the captured signals after injection and before extraction. The dark blue trace is the beam position sum signal, the red trace the fast beam loss monitor signal and the green trace the electron cloud signal.

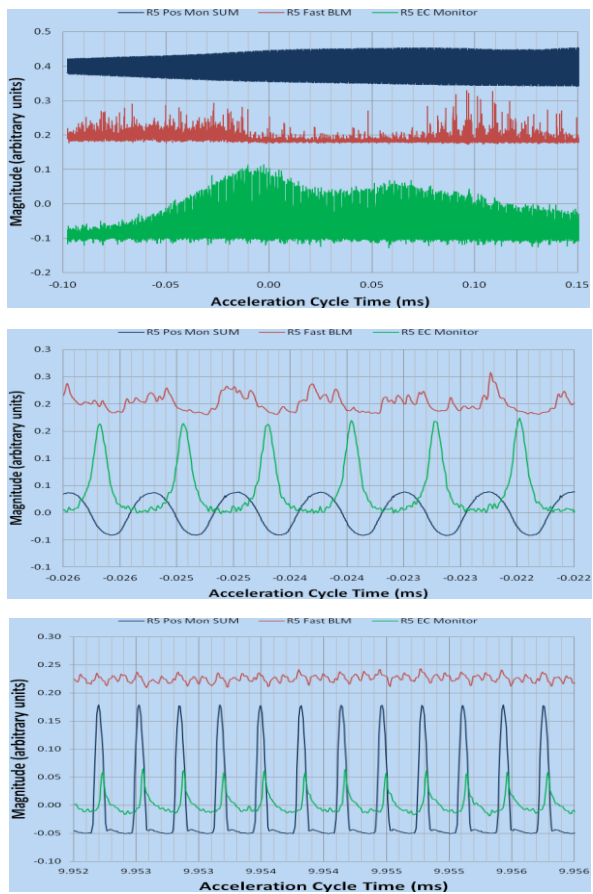

Figures 6, 7 and 8: EC signals captured at different stages during the acceleration cycle. Top: 250us of data taken after injection. Middle: 5us detail of the previous signal. Bottom: 5us of data taken before extraction.

Beam loses before extraction (Figure 8) are much smaller that after injection (Figure 7). The electron signal is also smaller at extraction, so background removal techniques had to be applied in order to obtain a clear signal

## Electron cloud, beam loses and beam position correlations

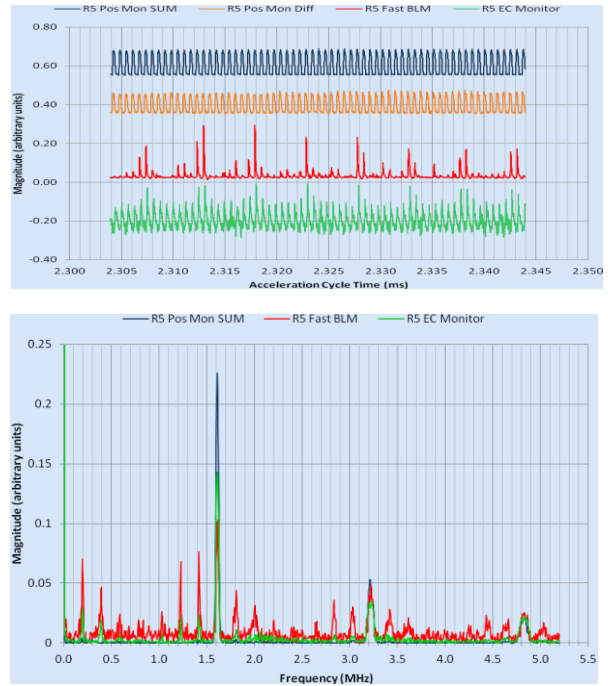

Figures 9 and 10: Oscillations observed on the EC and beam loss signals. Top: Time domain capture showing the correlation between both signals. Bottom: Frequency spectrum of the signals confirming the coincidence of the frequency components.

Measurements have shown a strong correlation between the beam loss signal and the electron cloud signal. Figure 9 shows an oscillation on the beam loss monitor signal (red trace) that correlates with a similar oscillation on the electron cloud signal (green trace).

The FFT graph (Figure 10) shows the coincidence of the sideband components of the signals. This oscillation on the beam loss signal does not appear on the horizontal beam position difference signal (orange trace). Additional measurements taken with an oscilloscope, connected to the differential signal of a nearby vertical position monitor, showed the same patterns on the oscillations and the FFT, suggesting that a high proportion of the electron cloud signal is a result of vertical instabilities which lead to an increase in secondary electron production.

# SUMMARY

With the new ISIS electron cloud monitor it has been possible to observe for the first time electron cloud signals generated during the acceleration cycle in the ISIS synchrotron. Initial results have shown the presence of a large electron cloud signal at the end of the injection stage and a strong correlation between measured beam loss and the electron cloud signal.

Vertical Beam instabilities as measured using position monitors display similar oscillation patterns as those observed in the beam loss and electron cloud signals. This suggests that it would be preferential to locate the electron cloud monitor in the vertical plane rather than the horizontal plane where there are no observed instabilities. A second electron cloud monitor will therefore be installed in the vertical plane in due course and these measurements repeated.

Possible improvements to the detector system are:

- Additional RFA detector (MCP based) on the vertical plane.
- Addition of the vertical position monitor difference signal to the data acquisition system in order to investigate the correlation between the existing vertical beam oscillations, beam loss and electron cloud signal.
- Installation of a D.C. amplifier setup for electron cloud build up measurements.

## ACKNOWLEDGEMENTS

We would like to thank Chris Warsop, Dean Adams, Bryan Jones, Hayley Smith, Di Wright, Sara Fisher and John Medland for all the help and support provided towards the EC studies.

## REFERENCES

[1] Ubaldo Iriso Ariz, "Electron Clouds in the Relativistic Heavy Ion Collider", Brookhaven National Laboratory, C-A/AP/#228 February 2006.
[2] A. Aleksandrov, V. Danilov, "Electron Accumulation in the PSR: An Attempt to Understand the Basic Facts", SNS/ORNL/AP Technical Memo, May 2000.
[3] K. Ohmi and F. Zimmermann, "Wake-field and fast head-tail instability caused by an electron cloud", Physical Review E, Volume 65, 016502, 2001 The American Physical Society.
[4] M. A. Furman, "A Preliminary Comparative Study of the Electron-Cloud Effect for the PSR, ISIS, and the ESS", LBNL-52872/CBP, Note 516, June 2003.
[5] G Bellodi, "Comparative Simulation Studies Of Electron Cloud Build-Up For ISIS And Future Upgrades", Proceedings of EPAC 2004.
[6] Chris Warsop, "Injection Upgrades For The Isis Synchrotron", Proc. of IPAC'10, Kyoto, Japan, 2010.
[7] S.J. Payne, "Beam diagnostics at ISIS", et al, HB2008 conference proceedings, Nashville, TN, 2008.
[8] U. Iriso, "Electron Cloud Observations at RHIC in Run-3 (2002/03)", Brookhaven National Laboratory, C-A/AP/#129, December 2003.